# Symmetry in magnetoelectric electromagnetism and magnetoelectric meta-atoms

E. O. Kamenetskii

School of Electrical and Computer Engineering,
Ben Gurion University of the Negev, Beer Sheva, Israel

**Abstract**
While in electromagnetism we have space-time symmetry, magnetoelectric (ME) processes are characterized by space-time symmetry breaking. Our goal is to show that quantum vacuum fields with both time reversal and space inversion symmetry breaking (so-called ME fields) can be observed in subwavelength regions of electromagnetic (EM) radiation. The sources of ME fields are ME meta-atoms – subwavelength elements made of magnetic insulators, where the dynamical ME behavior is due to modulations and topological coupling of magnetization and electric polarization. The interaction between EM and ME systems occurs through virtual structures: ME virtual photons. Experimentally, ME quantum vacuum states can be detected in a microwave cavity. The topic of these studies concerns fundamental aspects of ME quantum electrodynamics (MEQED).

## I. INTRODUCTION: ELECTROMAGNETISM AND MAGNETOELECTRICITY

When studying the relationship between electromagnetism and magnetoelectricity, one of the main questions is whether it is even possible to unite these two entities. Such unification is one of the key concepts in physics. This question seems to be the cornerstone of ME quantum electrodynamics. While in electromagnetism we have space-time symmetry, ME processes are characterized by space-time symmetry breakings. From this aspect, chirality (space symmetry breaking) and magnetism (time reversal symmetry breaking) can be considered as particular cases of magnetoelectricity. Neither electromagnetism nor magnetoelectricity are about the *interaction forces* between electric charges and electric currents. It is also important to note that the propagation of an EM wave is considered as a process of coupling between polar (e.g., *E*-field) and axial (e.g., *H*-field) vectors *in a region the size of a wavelength*. Meanwhile, in ME meta-atoms and ME materials, we observe the coupling between polar and axial vectors *in the subwavelength region* of EM radiation. This shows that ME polaritons, being a unification of electromagnetism and magnetoelectricity in the *dipole approximation*, are far from trivial systems.

In considering these issues, a brief historical excursion into some problems of electromagnetism and magnetoelectricity is necessary. Chronologically, the *understanding* of magnetism (and finally electromagnetism) *arose from the study of electricity*. Coulomb formulated his law in 1785. Galvani's experiments, and then Volta's invention of the battery in 1800, led to the emergence of sources of an electric current. At that time, it was said that there exist two types of electricity: *static electricity*, originated from electric charges (Coulomb) and *dynamic electricity*, produced by electric currents (Galvani, Volta). In 1820, Biot–Savart's law was formulated. Ørsted discovered that an electric current creates a magnetic field around it. This sparked the research into the relation between electricity and magnetism, as a result of which Ampère's and Faraday's laws were formulated. The Coulomb force was generalized to the Lorentz force. It was noted that the symmetry properties of magnetic field are different from the symmetry properties of magnetic field. While electric field is a true vector, magnetic field is a pseudovector. The direction of the vector does not depend on the choice of the coordinate system. The direction of the pseudovector is changed if we change the handedness of the coordinate system. This distinction between polar and axial vectors becomes important in understanding the effect of spatial symmetry on the solution to physical systems. Also, an important question is about time reversal symmetry: the electric field is invariant under time reversal, while the magnetic field flips sign. The linking between polar and axial vectors in EM waves is possible due to the use of two *curl operators* in Maxwell's equations: the curl of a polar vector is a pseudovector, and the curl of a pseudovector vector is a polar vector. EM wave propagation is a relativistic process. Using the scalar electric potential $\phi$ and the magnetic vector potential $A^{(m)}$, one expresses EM-wave equations in terms of *divergence and gradient operators* based on the Lorentz gauge [1].

Whereas we can assert that the *understanding* of magnetism (and eventually electromagnetism) *arose from electricity*, can we say that *understanding* of magnetoelectricity *arose from magnetism*? In 1957 Landau and Lifshitz showed that ME effects are possible in *magnetic* crystals [2]. The violation of the invariances under space reflection parity *P* and time inversion *T* are necessary conditions for the emergence of the ME effect. Landau's theory describes the ME effect in single-phase material through expansion of the free energy expression. This expression has an energy term which describes *ME coupling*:

$$W_{ME} = \alpha_{ij} E_i H_j, \qquad (1)$$

Where $\alpha_{ij}$ is a ME tensor [2]. For linear ME effect, *both T and P should be broken*, but the product *PT* is conserved: *by consecutive application of time reversal and spatial inversion we would return to the initial state*. The energy (1) should be a scalar, and thus the ME coefficient $\alpha_{ij}$ should be both *T* - and *P* -odd. Differentiation of equation of free energy with respect to electric and magnetic fields respectively leads to electric polarization and magnetization.

$$p_i = \varepsilon_0 \varepsilon_{ij} E_j + \alpha_{ij} H_j, \quad M_i = \mu_0 \mu_{ij} H_j + \alpha_{ij} E_j. \qquad (2)$$

The ME effect is manifested in the fact that electric polarization is linearly induced by a magnetic field, and magnetization is induced by an electric field.

Magnetoelectricity can be connected to various mechanisms of the relationship between magnetization and electric polarization in materials. In multiferroics, the symmetries associated with ferromagnetism and ferroelectricity are time reversal and space reversal symmetries. This is due to two kinds of ordering: for ferromagnetism, we have the spontaneous ordering of orbital and spin magnetic moments, for ferroelectricity, we have the spontaneous ordering of electric dipole moments. Currently, a new class of multiferroics, the so-called type-II multiferroics, is of great interest, in which the intrinsic ME coupling is due to electrical polarization induced by *spatially modulated spin structures*. This occurs when the ordering of electron spins breaks the spatial symmetry of the inversion. In multiferroics, both magnetic moments and electric dipole moments can be ordered, inducing robust macroscopic quantities such as magnetization and polarization [3, 4]. Most research on the ME effect has focused on spin degrees of freedom. Besides spin magnetization, orbital magnetization also contributes to the total magnetization that can be induced by an electric field. Such an orbital magnetization induced by an electric field is called the *orbital* ME effect [4 – 7].

The theory of magnetoelectricity with its numerous conceptual models is very controversial. The coupling between the electric and magnetic dipoles in multiferroics has little to do with Maxwell's electrodynamics. However, since the breaking of time reversal symmetry can occur not only due to the intrinsic magnetic ordering in the material but also in a dielectric moving in a constant electric field, it is argued that a linear ME effect can be observed. When macroscopic bodies move, we usually speak of velocities much less than the speed of light. For $v/c \ll 1$, we have constitutive relations for electrodynamics of moving dielectrics [2]:

$$\vec{D} = \varepsilon\vec{E} + \frac{\varepsilon\mu - 1}{c}\vec{v} \times \vec{H}$$
$$\vec{B} = \mu\vec{H} + \frac{\varepsilon\mu - 1}{c}\vec{E} \times \vec{v}$$
(3)

where the scalars $\varepsilon$ and $\mu$ are the material parameters in the rest (comoving) frame. It is presumed that for this structure with an *extrinsic* ME effect, there exists an *inertial* reference frame (in motion with respect to the laboratory frame) where the structure does not exhibit ME characteristics. Such a frame, which is not unique, is called a *quasi-rest frame* for the structure. It is also argued that magnetoelectricity is essentially a relativistic effect which is exhibited as is the relative motion of a medium represented as averaged distribution of electric and magnetic (uncoupled) dipoles [8, 9]. The theory of magnetoelectricity in moving systems was extended in Ref. [10]. It was shown that the constitutive relations for ME crystals can be formulated in a relativistic representation with a four-dimensional (4D) formalism. In this analysis, the authors touch upon important aspects related to the role of non-local effects.

It is worth noting here that it is generally accepted that Röntgen's experimental confirmation [11] that a dielectric sample *moving* through an electric field becomes magnetized was not only the first experiment of a ME effect, but also a confirmation that magnetoelectricity is essentially a *relativistic* effect. Röntgen did not call this effect the ME effect. The term "magnetoelectric" was introduced by Debye much later, in 1926. However, the issue concerns some more fundamental aspects. Röntgen's original experiment was carried out in a *rotational* (non-inertial) frame of reference. Therefore, an analysis based on Einstein's special theory of relativity in inertial frames of reference is inappropriate in this case. One could also say that this Röntgen's experiment is not related to the ME effect at all. It is simply an experimental confirmation of the phenomenon that a *circulating displacement current* causes a magnetic response. It is well known in modern research that circulating displacement currents in dielectric resonators exhibit a magnetic response [12–14].

Experimental observation of the ME effect in a moving dielectric sample raises many fundamental questions. In almost all practical cases, the sample operates in the nonrelativistic limit. For sufficient sensitivity of the experiment, the light confinement in the resonator due to the effects of interference should be used. This gives a strong-coupling regime of the light-matter interaction. When, for example, we consider a moving Fabry–Perot interferometer (FBI), many new characteristics must be taken into account. Due to the relativistic transformation of refractive index, the dielectric medium inside a moving FPI becomes anisotropic. We have the velocity dependence of FPI transmission and the relativistic Doppler shift associated with frame changes [15, 16].

When analyzing the interconnection of electromagnetism and magnetoelectricity, we also encounter a fundamental difference in the relationship between static and dynamic regimes in electromagnetism and magnetoelectricity. In electromagnetism, the static electric and magnetic fields are *not coupled*. In the dynamic regime, these fields are coupled due to Faraday's and Ampere's laws. The displacement current added by Maxwell allows us to predict electromagnetic waves. The electric and magnetic fields are considered as a single entity, the electromagnetic field, if we take into account the special theory of relativity. The main point here is the fact that in electromagnetism electrostatic and magnetostatic phenomena are considered to be a *quasistatic limit* of classical electrodynamics [1, 2]. In magnetoelectricity, the situation is completely different: the static electric and magnetic fields are *mutually coupled* by potential ME energy. How does this affect the dynamic regime? It is an open question whether static magnetoelectricity can be considered as a quasistatic limit of dynamic magnetoelectricity. We do not have any physical laws that unify static and dynamic magnetoelectricities. Nevertheless, our basic idea is that dynamic magnetoelectricity *arises from magnetism* similar to static magnetoelectricity. This means that we should not regard true dynamic magnetoelectricity without taking into account the potential ME energy. Considering dynamic magnetoelectricity as a relativistic effect, we assume that the unification of EM and ME phenomena is essentially due to relativity. In the following discussion, we will try to clarify our point of view. We will show that dynamic magnetoelectricity determines the properties of the ME quantum vacuum.

To consider the symmetry in ME electromagnetism we need to introduce the concept of a ME meta-atom. The ME meta-atom contains magnetic and dielectric subsystems. It is a local source of the so-called *ME field*. The ME vacuum field has both electric and magnetic field components but is not a field consisting of a combination of electric- and magnetic-dipole near fields. This is a relativistic field, with *rotating* electric- and magnetic-field components. In this paper, we show that in the dynamic ME regime, the near field in vacuum has not only ME energy, but also spin and orbital angular momenta. We consider the ME meta-atom as a quantum emitter with *ME-energy eigenstates*. The interaction of EM radiation with ME meta-atoms is carried out through virtual ME photons.

It is worth noting that that the ME field is an *emergent field*. It arises from the collective behavior of magnetic and dielectric subsystems, rather than being fundamental constituents of these systems themselves. Unique macroscopic property of ME field appears when many microscopic magnetic- and electric-dipole components interact in a specific way.

## II. MAGNETOELECTRICITY AND ME FIELDS

**A. Static magnetoelectric interactions**

Let us begin our analysis with a purely static problem. Consider static interactions between small spheres made of different materials. The structure of such spheres is shown in Fig. 1. The dipole-dipole interactions arise from the combination of permanent or induced dipoles. When electric dipole moments are induced in small dielectric spheres, which are located close to each other, the entire system (inside the spheres and in vacuum region outside) is characterized by electric energy. When magnetic dipole moments are induced in small magnetic spheres located close to each other, the system is characterized by magnetic energy. Now let *both* electric and magnetic dipole moments are induced in two ME-material spheres that are also located close to each other. When the electric and magnetic moments are coupled by free energy expressed by Eq. (1), it can be said that the entire system of ME spheres must be characterized by a combination of electric energy, magnetic energy, and ME energy. These *three* types of energy are observed inside the spheres. But can we observe all these energies in the nearby *vacuum region* outside the ME samples?

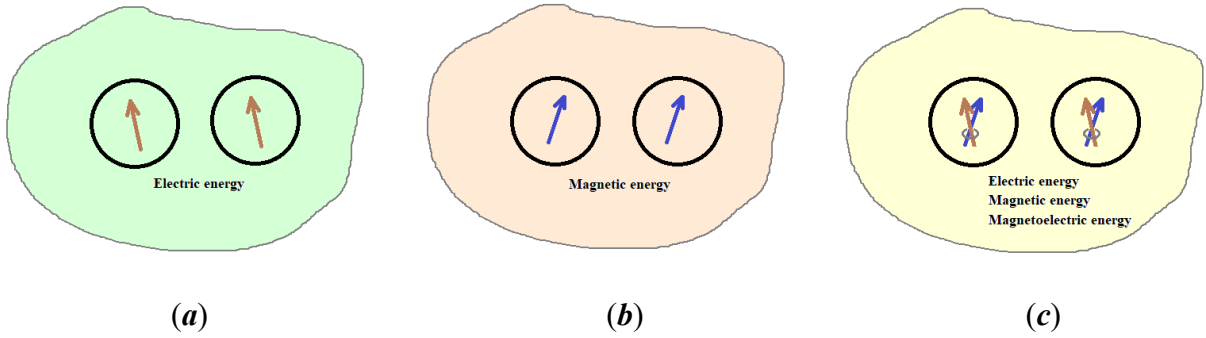

(*a*)          (*b*)          (*c*)

**Fig. 1**. Dipole–dipole interactions between two permanent or induced dipoles are characterized by different types of energies both inside the spheres and in the vacuum regions outside. (*a*) For two dielectric spheres, we have *electric field energy*. (*b*) For two magnetic spheres, there is *magnetic field energy*. (*c*) For two ME-material spheres, the fields should be characterized by electric energy, magnetic energy, and *ME energy*.

The constitutive relations in the static ME regime provide a basis for understanding how the internal polarization and magnetization of a material respond to applied static electric and magnetic fields, taking into account ME coupling effects. For ME media, constitutive relations in a static regime are:

$$\vec{D} = \ddot{\varepsilon}\vec{E} + \ddot{\alpha}\vec{H}, \quad \vec{B} = \ddot{\mu}\vec{H} + \ddot{\alpha}\vec{E}. \tag{4}$$

It should be noted here that the equality $\ddot{\alpha}_{em} = \ddot{\alpha}_{me} = \ddot{\alpha}$ holds only in the absence of dissipation and dispersion, which describes the low frequency responses [17]. In multiferroics, both magnetic moments and electric dipole moments can be ordered, giving rise to robust macroscopic quantities such as magnetization and polarization. At the same time, in electrodynamics, boundary conditions are imposed on electric and magnetic fields, not on electric polarization and magnetization. From Eqs. (4), the electromagnetic boundary conditions for normal and tangential fields on the surface of the ME (number 1) and vacuum (number 2) regions are as follows:

$$\left(\vec{\vec{\varepsilon}}\cdot\vec{E}_1\right)_n + \left(\vec{\vec{\alpha}}\cdot\vec{H}_1\right)_n = \varepsilon_0\left(\vec{E}_2\right)_n$$

$$\left(\vec{E}_1\right)_t = \left(\vec{E}_2\right)_t$$

$$\left(\vec{\vec{\mu}}\cdot\vec{H}_1\right)_n + \left(\vec{\vec{\alpha}}\cdot\vec{E}_1\right)_n = \mu_0\left(\vec{H}_2\right)_n$$

$$\left(\vec{H}_1\right)_t = \left(\vec{H}_2\right)_t \qquad (5)$$

Compared to dielectric and magnetic samples, the ME sample is characterized by *both* electric and magnetic surface charges.

We know that the determination of the internal material structure can be realized by using electrostatic or magnetostatic methods. The example is force microscopy. The force microscopy images the stray electric or magnetic field around a sample. The stray field itself is caused by electric and magnetic volume and surface charges in the sample, which on their turn find their origin in the non-uniform distribution of the electric polarization or magnetization. The results are obtained by solving the Laplace equations for electrostatic potential, $\nabla^2\phi = 0$, or for magnetostatic potential $\nabla^2\psi = 0$. Currently the ME force microscopy is available. The basic principle is to use magnetic force microscopy to analyze the electric field-induced magnetization [18, 19]. Similarly, one can propose to use electric force microscopy to analyze the magnetic field-induced electric polarization. Such techniques for detecting local ME responses are not ME energy detection. Both the magnetostatic (the magnetic field *T*-odd) and electrostatic (the electric field is *P*-odd), laws do not imply any *mutual coupling* between point electric and magnetic dipoles in vacuum. No differential equation of connection of static electric and magnetic potentials ($\phi \leftrightarrow \psi$) exists. There is no "ME scalar potential".

The abrupt transition of the constitutive parameters at the sample surfaces is not an abrupt transition of the fields. The fact that the vacuum electric and magnetic fields will somehow be modified by the *intrinsic* ME effect into the samples does not give us the answer we need. Since in statics ME energy is a purely intrinsic notion, we do not observe violations of the *T* and *P* symmetries for near-fields *in vacuum*. Vacuum does not bear traces of the ME effect in statics. Can we measure ME energy with the use of a field with both *T* - and *P* -odd in vacuum? An example of such a field is an axion field [20]. The violation of both *T*- and *P*-symmetry means that the laws of physics are not the same when considering a given process and the process in its time-reversed analogue and mirror image. If *T* and *P* are violated, C (charge conjugation) must also be violated to maintain the *CPT* symmetry. So, if a system violates both *T* and *P* symmetries, then it must also violate CP symmetry. In essence, the violation of *T* and *P* in a ME field provides evidence for *CP* violation.

The symmetry-breaking properties of ME material should leave a characteristic imprint in virtual photons of quantum vacuum fluctuations near a sample. Probing quantum atmospheres of axion fields might offer a way to identify new types of symmetry breaking effects [21]. The question now arises: is it possible to realize these effects with both *T*- and *P*-odd vacuum fields in the case of *dynamic magnetoelectricity*?

**B. Dynamic magnetoelectricity and ME energy**

In the non-time-varying regime, the electric and magnetic fields are not interconnected in a vacuum near-field region of a sample with the *intrinsic* magnetoelectricity. So, no vacuum fields with space and time symmetry breakings – the ME fields – can be observed in the case of static magnetoelectricity. The above analysis indicates that for EM radiation each small sample with the static ME effect is viewed not as the *P*- and *T*-breaking singularities in a *deep subwavelength* region of EM radiation, but as a specific region with some combination of electric and magnetic dipoles. Even taking into account the ME-coupling coefficients inside the ME structure, the use of standard EM boundary conditions does not show the ME property (and therefore ME energy) of near fields in vacuum. Is it becoming obvious that the structure of magnetoelectric electromagnetism – the structure with the united ME-EM field – must be based not on the effects of static magnetoelectricity, but on the effects of dynamic magnetoelectricity?

Dynamic magnetoelectricity is the generation of electrical polarization by means of magnetization changing in space and time. These are dynamic properties of resonantly oscillating magnetizations and *magnetically induced* polarizations. In Ref. [22], electric activity of spin waves was considered. It was predicted that such spin wave can carry, besides magnetization, also an electric dipole perpendicular both to magnetization and to the magnon wavevector. Dynamic magnetoelectricity in helical magnets and magnetic skyrmions was considered in Ref. [23 – 27]. These are ME effects of coupling topological spin textures and electric polarization via *relativistic spin-orbit interaction*. The goal of these works was to study the unique EM effects due to the resonances of collective ME modes. It is worth noting that in these studies, only the energy of the electric and magnetic fields is considered, but *not the energy of the ME coupling*. In different theoretical models of interaction of the ME structure with photons (such as dipolar cross-coupling between the induced polarization and magnetization [24, 25], magnon polaritons [26], and Zeeman and Stark interaction [27]), the *effects related to ME energy are not even discussed*.

In the dynamic regime, the observation of ME characteristics in vacuum EM fluctuations is the subject of much discussion. Quantum fluctuations near the ME material with violation of the *PT* discrete symmetries will produce a sort of *PT violating atmosphere* [21]. It is argued that this atmosphere induces new kinds of "Casimir" forces on bodies near the material. There are claims that the vacuum can impart momentum asymmetrically on ME structures. Asymmetric momentum transfer arises from the ME structure since it breaks the temporal and spatial symmetries of electromagnetic modes. The possibility of extracting *linear momentum* from vacuum was discussed by Feigel [28]. Feigel argued that the momentum of vacuum zero fluctuations can occur only in a structure with *PT*-symmetry breakings. The main idea is to suggest a new quantum mechanical effect, namely the extraction of momentum from the electromagnetic vacuum oscillations. In the proposed effect, linear momentum is extracted from a vacuum field. This is different from the case of the Casimir effect, in which energy is extracted from a vacuum field [29]. It was argued that rotating ME particles can generate changes in momentum of zero-point fluctuations, which result in the "self-propulsion" in quantum vacuum. The "self-propulsion" in quantum vacuum requires mechanical back-action from ME particle. To provide this, the ME particle should be a propellor-like device [30]. Thus, the fundamental question of extracting *angular momentum* from the vacuum arises.

For ME medium in a dynamic regime, we have time-dispersive constitutive relations:

$$\vec{D}(\omega) = \ddot{\varepsilon}(\omega)\vec{E} + \ddot{\xi}(\omega)\vec{H}, \qquad \vec{B}(\omega) = \ddot{\zeta}(\omega)\vec{E} + \ddot{\mu}(\omega)\vec{H}. \tag{6}$$

While considering the interaction of dynamic ME materials with EM radiation, we assume that materials are *local*. This means that the coupling between electric and magnetic fields occurs in the subwavelength region $l$ of the EM radiation:

$$l \ll \lambda_{EM}, \tag{7}$$

where $\lambda_{EM}$ is the size of the wavelength in the medium. The use of constitutive relations (6) is based on the assumption of a continuum model of the medium. This model can be applied to metamaterial arrays composed of ME structural elements. However, the question arises about the nature of the fields by which these ME elements interact with each other. We know that in metamaterial arrays composed of dielectric or magnetic structural elements, the interacting fields are *static electric or magnetic fields*, respectively. At the same time, based on our above argumentation and known experiments, we can say that in ME metamaterials, the structural elements do not interact with *static ME fields* – the fields with both $T$- and $P$-odd symmetry.

The problem associated with the nature of the ME field and ME energy certainly arises when analyzing the interaction of the ME medium with EM radiation using energy relations [31]. To derive the energy balance equation in temporary dispersive media, one must use the regime of propagation of *quasi-monochromatic* EM waves [2]. For a ME medium, such quasi-monochromatic behavior was considered in Refs. [32 – 34]. The fields are expressed as $\vec{E} = \vec{E}_m(t,\vec{r})\, e^{i(\omega t - \vec{k}\cdot\vec{r})}$ and $\vec{H} = \vec{H}_m(t,\vec{r})\, e^{i(\omega t - \vec{k}\cdot\vec{r})}$, where complex amplitudes $\vec{E}_m(t,\vec{r})$ and $\vec{H}_m(t,\vec{r})$ are time and space smooth-fluctuation functions. The average energy density is expressed as

$$\langle W \rangle = \frac{1}{4}\left\{ \frac{\partial(\omega \varepsilon_{ij}^h)}{\partial \omega} E_i^* E_j + \frac{\partial(\omega \mu_{ij}^h)}{\partial \omega} H_i^* H_j + \frac{\partial\left[\omega(\zeta_{ij}^h + \xi_{ij}^h)\right]}{\partial \omega}\left(H_i^* E_j\right)^h + \frac{\partial\left[\omega(\zeta_{ij}^{ah} - \xi_{ij}^{ah})\right]}{\partial \omega}\left(H_i^* E_j\right)^{ah} \right\}, \tag{8}$$

where superscripts $h$ and $ah$ in Eq. (8) denote, respectively, the Hermitian and anti-Hermitian parts of tensors of the second rank. For a lossless medium, tensors $\ddot{\varepsilon}$ and $\ddot{\mu}$ are Hermitian and $\ddot{\xi}^h = \ddot{\zeta}^h$, $\ddot{\xi}^{ah} = -\ddot{\zeta}^{ah}$. The electric-field energy density $W_E$ and the magnetic-field energy density $W_M$ are expressed, respectively, by the first two terms on the right-hand side of Eq. (8). The ME energy density $W_{ME}$ is expressed by the last two terms on the right-hand side of Eq. (8). While the densities $W_E$ and $W_M$ are the energies of dielectric and magnetic materials in the *static* electric and magnetic fields, respectively [1, 2], the ME energy density $W_{ME}$ in Eq. (8) does not corresponds to the *static* ME energy expressed by Eq. (1). From Eq. (8) we see that in ME electrodynamics, the *material structure and the field structure are inextricably linked*. In general, ME energy density $W_{ME}$ should be a complex quantity. Moreover, in Ref. [32] it was shown that the Poynting-theorem continuity equation can be valid only when certain constraints are imposed to the slowly time-varying amplitudes of the fields. For the *i*-th and *j*-th components of vectors of the electric and magnetic field amplitudes, $\vec{E}_m$ and $\vec{H}_m$, these constraints are expressed as $E^*_{m_i}(t)\frac{\partial H_{m_j}(t)}{\partial t} = H_{m_j}(t)\frac{\partial E^*_{m_i}(t)}{\partial t}$. In a general form, these constraints can be written as $E_{m_i}(t) = T_{ij} H^*_{m_j}(t)$. The matrix $[T]$ is a field-polarization matrix. The components of matrix $[T]$ are pseudoscalar complex quantities consisting of a constant and second-harmonic oscillating part [32 – 34]. Since subwavelength fluctuations in the intensity of EM excitation in a ME medium arise only at certain ratios of the amplitudes of the electric and magnetic fields, energy conversion between average stored energies $\langle W_E \rangle$, $\langle W_M \rangle$, and $\langle W_{ME} \rangle$ can occur through a certain *circular process of energy exchange*. This implies the presence of local *power-flow circulation*, which can be carried out with *synchronous rotation* in time of both complex-amplitude vectors $E_{m_i}$ and $H_{m_j}$. This means that the process of propagation of EM waves in a ME media is accompanied by *power-flow vortices in subwavelength domains*.

The energy balance equation for EM waves propagating in the ME medium, showing the unique effect of power-flow circulations in local regions, reveals the fact that such a subwavelength circulation process of energy should also occur in small ME resonators – ME meta-atoms. In this meta-atom, we have a ME "trion"– a localized (subwavelength) resonant excitation with energies of three subsystems (electric, magnetic, and magnetoelectric). The energy states of ME "trions" can be split in an applied magnetic field. For two opposite directions of a normal bias field, there are two types of ME "trions" with different chirality. Since the temporal inversion symmetry is broken, we can observe the right- and left-hand power-flow vortices [34]. In contrast to the well-known effect of thermal equilibrium of persistent Poynting flux in non-reciprocal many-body EM systems [35 – 37], we observe persistent energy currents at a non-equilibrium state in EM subwavelength regions with *quasistatic oscillations*. There are rotational energy flows, both clockwise and counterclockwise.

The concept of ME quantum vacuum involves the study of vacuum states of quantized ME fields arising from the oscillation spectra of ME resonators. Fundamental studies of the interaction of a resonant point ME scatterer with EM radiation constitutes a field of research called ME quantum electrodynamics. The main aspects of the physics of ME meta-atoms should be related to magnetism, relativity theory, and topology.

## III. QUASISTATIC OSCILLATIONS AND ME FIELDS

In condensed matter electrodynamics, the concept of locality implies the existence of a *quasistatic limit*. Usually, this means that the internal energies of materials in the dynamic regime are defined in the same sense as in the static regime, i.e. for $\omega = 0$ [2]. In the case of ME materials, we should consider another form of the quasistatic limit. If we accept the fact that the internal ME energy must be defined at $\omega \neq 0$, the quasistatic limit appears at $|\vec{k}_{EM}| \to 0$. It implies the presence of local (subwavelength) regions with *quasistatic oscillations*. These quasistatic oscillations have wave numbers $|\vec{k}_{QS}|$ significantly exceeding the EM wave number $|\vec{k}_{EM}|$: $|\vec{k}_{QS}| \gg |\vec{k}_{EM}|$.

In the dynamic regime supported by quasistatic oscillations, no electromagnetic retardation effects are assumed. To observe vacuum ME fields – fields with both *T* and *P* violations – we need to obtain a local source of dynamic magnetoelectricity. Such a ME meta-atom should be a 3D structure in the *subwavelength region* of EM radiation. The interaction of a small ME resonator with EM radiation must be of a quantum nature. It means that the meta-atom has discrete energy levels that arise from the wave behavior of the intrinsic ME oscillations. For these ME oscillations in meta-atoms, we must observe *energy circulation*.

In classical Maxwell electrodynamics, electricity and magnetism are two phenomena related by duality symmetry. There are two curl operators which couple the polar and axial fields vectors. We now formally introduce *two systems* of differential equations, each using *only one* curl operator. Assuming initially that these systems are not coupled, we will describe two separate *quasistatic structures*. The first structure is represented by equations

$$\nabla \cdot \vec{B} = 0 \tag{9}$$

$$\vec{\nabla} \times \vec{H} = 0 \tag{10}$$

$$\vec{\nabla} \times \vec{E} = -\frac{\partial \vec{B}}{\partial t} \tag{11}$$

In this case, electric displacement current is excluded, $\frac{\partial \vec{D}}{\partial t} = 0$. Magnetic field is quasistatic: $\vec{H} = -\vec{\nabla}\psi$. This system of equations describes *magnetostatic (MS) oscillations* in magnetic insulator samples.

The second structure is described by equations

$$\nabla \cdot \vec{D} = 0 \tag{12}$$

$$\vec{\nabla} \times \vec{E} = 0 \tag{13}$$

$$\vec{\nabla} \times \vec{H} = \frac{\partial \vec{D}}{\partial t} \tag{14}$$

Here magnetic displacement current is excluded, $\frac{\partial \vec{B}}{\partial t} = 0$. We have quasistatic electric field: $\vec{E} = -\vec{\nabla}\phi$. This system of equations can characterize *electrostatic (ES) oscillations* in dielectric samples.

MS resonances are observed in a small sample of magnetic insulator material with strong temporal dispersion of permeability, $\vec{B} = \vec{\mu}(\omega) \cdot \vec{\nabla}\psi$. From Eqs. (9), (10) we have wave equation for the MS oscillations

$$\vec{\nabla} \cdot \left( \overleftrightarrow{\mu} \cdot \vec{\nabla} \psi \right) = 0, \tag{15}$$

where $\overleftrightarrow{\mu}(\omega)$ is the permeability tensor. MS resonances are magnetic dipole oscillations. In this case, we neglect the time variation of the energy of the electric field as compared to the time variation of the energy of the magnetic field. In analyses of MS-wave problems in a magnetic insulator, Eq. (15) is known as Walker's equation [38]. It is worth noting that in the literature, where the problems on the long-range magnetization oscillations and waves in ferrite samples are studied, one can see that in the spectral problems formulated exceptionally for the MS-potential wave function $\psi$, no consideration is given to the electric fields arising from Eq. (11), the Faraday law [1, 39 – 42]. However, when studying ME coupling effects in 3D-confined subwavelength ferrite samples, the role of electric fields is fundamental [43].

With use Eq. (11) we can show that if a sample has isotropic permittivity, there is $\frac{\partial^2 \vec{B}}{\partial t^2} = 0$. This means that it is physically incorrect to consider the curl electric field in the MS resonances. Let us assume, however, that the magnetic-dipole dynamics in a magnetic insulator sample is accompanied by an *induced* electric polarization which is characterized by temporally dispersive permittivity tensor $\overleftrightarrow{\varepsilon}_{ind}(\omega)$. In such an assumption with using $\vec{D}_{ind} = \overleftrightarrow{\varepsilon}_{ind}(\omega) \vec{E}$, we have $\frac{\partial^2 \vec{B}}{\partial t^2} \neq 0$. So, the curl electric field can be observed, and certain magnetic-electric interrelation becomes restored in the MS resonances in a magnetic-insulator sample. Electric polarization can be induced due to a topological effect associated with *chiral magnetic currents* at the boundary of a sample. These are topologically protective edge currents in a *non-simply connected domain*. While analyzing what is a suitable geometrical shape of a 3D confined subwavelength ferrite sample where topological edge currents can occur, we should not consider the spherical sample because the sphere is simply connected; and thus, every current loop can be contracted on the surface to a point. Samples should also not be in the shape of an ellipsoid and an infinite cylinder. At the same time, a quasi-2D disk turns out to be the suitable sample shape. When the current loop is on the lateral surface of such a disk, we have a non-simply connected domain with topologically protective edge currents.

The system of equations (12) – (14) has the same form as the system of equations (9) – (11). Therefore, for ES oscillations, the same analysis can be formally applied as for MS oscillations. For a dielectric material with strong temporal dispersion of permittivity $\vec{D} = \overleftrightarrow{\varepsilon}(\omega) \cdot \vec{\nabla} \phi$, from Eqs. (12) and (13) we have a wave equation for ES oscillations

$$\vec{\nabla} \cdot \left( \overleftrightarrow{\varepsilon} \cdot \vec{\nabla} \phi \right) = 0. \tag{16}$$

ES resonances are electric dipole oscillations. In this case, we neglect the time variations of the energy of the magnetic field as compared to the time variations of the energy of the electric field. In ES oscillations, the electric-magnetic interrelation is restored taking into account the curl magnetic field of Eq. (14). In order to have $\frac{\partial^2 \vec{D}}{\partial t^2} \neq 0$, the electric-dipole dynamics in a dielectric sample should be accompanied with the *induced magnetization* characterized by temporal-dispersion permeability $\ddot{\mu}_{ind}(\omega)$. In such an assumption, $\vec{B}_{ind} = \ddot{\mu}_{ind}(\omega)\vec{H}$. So, the curl magnetic field can be observed, and certain magnetic-electric interrelation becomes restored in the ES resonances. Magnetization can be induced due to a topological effect associated with *chiral electric currents* at the boundary of a sample.

Equations (9) – (11) and (12) – (14) together with the constitutive relations of the induced anisotropic permittivity and permeability, $\vec{D}_{ind} = \ddot{\varepsilon}_{ind}(\omega)\vec{E}$ and $\vec{B}_{ind} = \ddot{\mu}_{ind}(\omega)\vec{H}$, may constitute a symmetric mathematical structure that we characterize as *ME duality*. ME duality focuses on scenarios involving time-varying electric and magnetic fields in specific materials and specific topological structures.

Following our basic idea that magnetoelectricity arises from magnetism, we argue that the ME meta-atom is a magnetic-insulator structure with *eigen* MS resonances and *induced* ES oscillations. The systems of equations (9) – (11) and (12) – (14), being connected in small ferrite samples of a *certain geometry*, can represent solutions for the fields of the ME structure. In these meta-atoms, the dynamic ME behavior is due to modulations and topological coupling of magnetization and electric polarization. ME meta-atoms are sources of ME fields. The dynamic ME fields manifest themselves as *relativistic fields* of coupled quasistatic, MS and ES, resonances [43].

The main role of coupling of MS and ES resonance belongs to magnon spin-orbit interaction in a quasi-2D disk of magnetic insulators [44]. In this case, we have a quasi-rotating ME system: A system that exhibits rotational characteristics (being analyzed within a rotating reference frame) but it's not a real rotation. Similar to Barnett's effect [45] of magnetization by rotation, we observe electric polarization due to orbital angular momentum of magnetic-dipolar oscillations (see experiments in Ref. [46]). In such a meta-atom, the effect of ME coupling vanishes in a rotating reference frame. The ME properties are manifested in the lab frame, but not in a rotating frame, making them appear "quasi-rotating" from the perspective of the lab frame. There is a time-dependent ME structure: The term "quasi-rotating" in this context refers to how the ME properties over time demonstrate a periodic pattern of changes.

It is worth noting that the MS and ES resonances described here are not related to magnon-ferron coupling which refers to the interaction between magnons (spin waves) and ferrons (excitations of ferroelectric polarization) in multiferroic materials [47].

**IV. ME RESONANCES OF META-ATOMS**

In a sample made of a magnetic insulator, an interaction between the ferromagnetic order subsystem and the electric polarization subsystem can occur. Similar to type II multiferroics, *the intrinsic ME coupling in the ferrite disk is due to the electric polarization caused by spatially and temporally modulated spin structures*, but the physical mechanism is completely different. In ME meta-atoms, the mesoscopic effect of dynamic magnetoelectricity arises from the topologically coupled MS and ES resonances. Both MS and ES oscillations are observed with quantum confinement effects for scalar *wave functions* $\psi(\vec{r},t)$ and $\phi(\vec{r},t)$ in a small sample localized in the subwavelength region of EM radiation. The term "ME coupled" implies that the MS and ES states are not isolated but rather interact with each other due to topologically induced edge currents. In general, this interaction is expressed by terms in the Hamiltonian that relates to the ME states: $\langle \psi | \mathrm{H}_{ME} | \phi \rangle$. The ME coupled states can be represented using a tensor product of the individual, MS and ES, state spaces. In the ME resonance states, a violation of both spatial and temporal inversion symmetry occurs. ME duality involves a symmetry relationship between time-varying electric and magnetic fields that is distinct from EM fields. Any EM retardation effects are disregarded.

In a magnetic sample, MS oscillations are magnetic-dipolar modes (MDMs) [2, 38 – 42]. When analyzing these modes in a quasi-2D ferrite disk, two types of spectral solutions should be considered. These solutions, called $G$ and $L$ modes [44], arise from the fact that for the scalar wave functions $\psi(\vec{r},t)$, the second-order equation (15), on the one hand, and the system of two first-order equation (9) and (10), on the other hand, have *different boundary conditions*, respectively. For $G$ modes, we define the energy eigenstates of MS oscillations based on the Schrödinger-like equation for wave function $\psi(\vec{r},t)$ using the Neumann-Dirichlet (ND) boundary conditions. In the case of $L$ modes, we consider normalization to the power-flow density using the EM boundary conditions (continuity of the $\vec{H}$- and $\vec{B}$-field components). Magnetic chiral currents induced at the lateral boundary of the disk are associated with the differences between the ND and EM boundary conditions. The mesoscopic basis for the surface magnetic current at the interface between gyrotropic and nongyrotropic media is provided by the Berry mechanism. A quasi-2D disk is a non-simply connected domain with topologically protective magnetic edge currents. This non-simply connected region with the current loop is obviously associated with the electric-field flux. Such a description of polarization may be regarded as a multi-valued quantity [48 – 50].

Due to circulation of a chiral magnetic current on a lateral surface of a ferrite disk, electric charges appear on the top and bottom planes of the ferrite disk. This induces a normal electric-field gradient. As a result, we have the *electric-quadrupole precession caused by the magnetization dynamics*. The magnetic-dipole and electric-quadrupole resonances are well known in nuclear physics [51]. The electric field gradient arises from the inhomogeneous distribution of charges. Electric quadrupole precession, in the context of nuclear magnetic resonance (NMR) refers to the precession of nucleus's quadrupole moment in an electric field gradient. This precession is distinct from regular Larmor precession, which is caused by a magnetic field. In our case, we have a unique effect of coupling these resonances. The spectral analysis for ES scalar wave functions $\phi$ in a quasi-2D ferrite disk will be like the spectral analysis for MS scalar wave functions $\psi$. Similar to magnetic oscillations, for electric oscillations, the differences between the ND and EM boundary conditions will lead to the appearance of *electric chiral currents* at the lateral boundary of the disk sample. Because of this topologically protective electric edge currents the magnetic-field flux is observed.

The model predicts the topological ME effect, where an edge current of orbital magnetization generates a topological contribution to electric polarization and an edge current of orbital polarization generates a topological contribution to magnetization. Topological currents provide us with the possibility to have chiral rotational symmetry by the turn over a regular-coordinate angle $2\pi$ at the $\pi$-shift of a dynamic phase of the external EM field. The frequency of orbital rotation must be twice the EM wave frequency $\omega$. In the coordinate frame of orbitally driven field patterns, the lines of the electric field $\vec{E}$ as well the lines of the polarization $\vec{p}$ are "frozen" in the lines of magnetization $\vec{m}$. It means that there are no time variations of vectors $\vec{E}$ and $\vec{p}$ with respect to vector $\vec{m}$. It is worth noting that for $L$ modes, in a relativistically rotating frame of reference we observe a reduced magnetic energy compared to the laboratory frame of reference. This leads to the appearance of quantized electric energy. In the laboratory frame, we observe spin and orbital momenta both for the polarization and magnetization vectors. In near-field vacuum region, there are power flow vortices and non-zero ME energy.

The theory of the spectral properties of ME oscillations in a ferrite disk resonator is published in [46, 52 – 58]. Several illustrations related to this theory, emphasizing the main aspects necessary for the present study, are shown in Figs. 2 – 5.

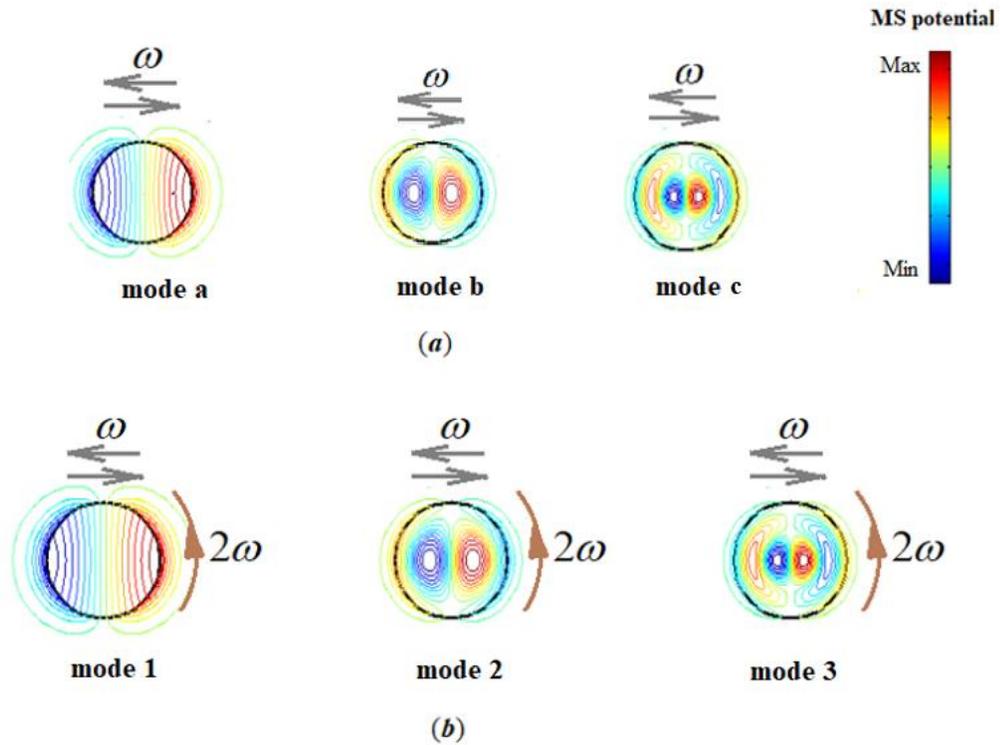

Fig. 2. MS-potential distributions for $G$ and $L$ modes in a quasi-2D ferrite disk viewed in the laboratory coordinate system. For $G$ modes, there is a collinear magnetic system, where spins align in parallel or antiparallel configurations. For $L$ modes, non-collinear arrangements exhibit spatially varying spin orientations that give rise to topologically non-trivial spin textures due to chiral rotation in systems lacking inversion symmetry. Lacking inversion symmetry is due to the involvement of the electric-dipole-polarization subsystem (see Fig. 3 below). (*a*) The first three $G$ modes. (*b*) The first three $L$ modes.

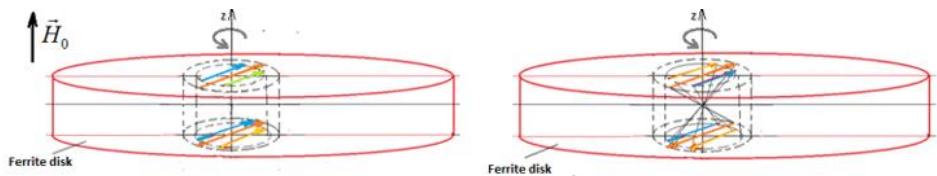

(*a*)    (*b*)

Fig. 3. The figure shows how local magnetic and electric moments inside the ferrite disk align and rotate in the *L* mode. In the laboratory coordinate system, we see magnetic-dipole (*a*) and electric-quadrupole (*b*) structures rotating at a frequency twice the frequence of the microwave signal. In a rotating coordinate system, the lines of polarization $\vec{p}$ are "frozen" in the lines of magnetization $\vec{m}$. The inversion symmetry is broken due to the electric-quadrupole structure (Ref. [43]). In magnetic resonance, the rotating frame simplifies the description of spin dynamics by effectively "freezing out" the time evolution of the spins due to the applied radiofrequency field.

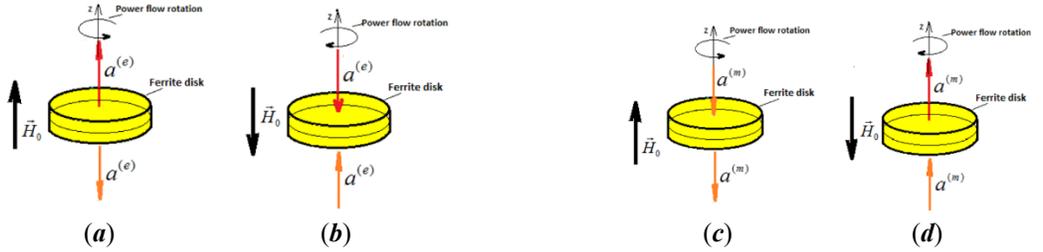

(*a*)    (*b*)    (*c*)    (*d*)

Fig. 4. Rotation may induce fictitious electric and magnetic charges. We observe electric and magnetic fluxes at the *L*-mode resonances. (*a*), (*b*) In the magnetic subsystem of the ferrite-disk resonator, localized distribution of an *edge magnetic current* is viewed as an *electric flux*. The electric moment $\vec{a}^{(e)}$ is considered as the density of the electric flux. This structure with opposite polarization directions on the top and bottom planes of a ferrite disk leads to appearance of the electric-field gradient along *z* axis. This electric-field gradient acts like a torque on the electric-quadrupole moment, causing it to precess. (*c*), (*d*) Localized distribution of an *edge electric current* in the electric subsystem is viewed as a dipole magnetic field at a large distance. The magnetic moment $\vec{a}^{(m)}$ is considered as the density of the *magnetic flux*. The arrows indicate the *z*-direction of the magnetic field flow at the center of the disk. (Ref. [43]).

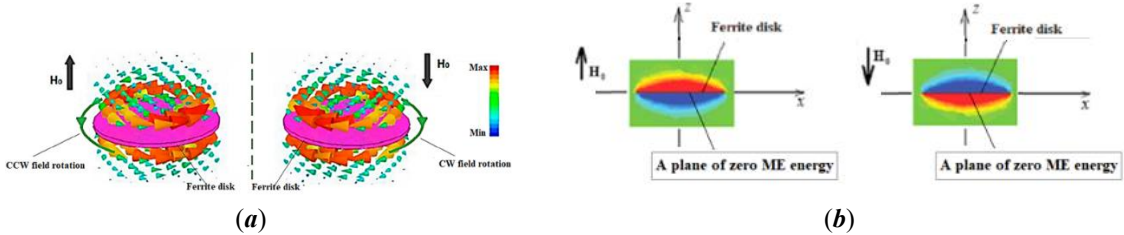

(*a*)    (*b*)

Fig. 5. Vacuum near-field region for the first *L* mode. (*a*) Power-flow rotation. (*b*) Positive (red colored) and negative (blue colored) ME energy (Ref. [58]).

Modes a, b, c shown in Fig. 2 are ground states. These are quasi-rest-frame structures. "Quasi-rest frame structures" refers to analyzing a system or object as if it were at rest, even though it might be in motion or undergoing acceleration, from the perspective of a non-inertial or accelerating frame of reference. This approach allows for simplification of complex dynamic problems by considering fictitious forces that arise due to the acceleration of the reference frame. Modes 1, 2, 3 in Fig. 2 are *metastable states*. *G* modes are MDMs. There is no ME coupling for *G* modes. But *L* modes are ME modes.

The main role of coupling of ME and ES resonance belongs to magnon spin-orbit interaction in a quasi-2D disk of magnetic insulators. In this case, we have a quasi-rotating ME system: A system that exhibits rotational characteristics (being analyzed within a rotating reference frame) but it's not a real rotation. In such a meta-atom, the effect of ME coupling vanishes in a rotating reference frame. The term "quasi-rotating" in this context refers to how the ME properties over time demonstrate a periodic pattern of changes.

## V. ME QUANTUM STATES IN A CAVITY

Before proceeding to the consideration of ME quantum vacuum states, it is appropriate to briefly dwell on the main characteristics of the oscillation modes in the quasi-2D ferrite-disk ME meta-atom. For L-modes we have a chiral rotation. In such a rotation the inversion symmetry is broken due to the electric quadrupole structure. In this case the magnetic system, being collinear in the rotational coordinate system, is represented as non-collinear dipole configurations in the laboratory coordinate system. This leads to topologically non-trivial textures of the field vectors in the laboratory coordinate system. ME fields, the fields observed in vacuum in close proximity to the L-mode ferrite disk, are non-collinear near fields. In this field structure, the electric and magnetic field vectors are directed in different directions, creating complex field configurations. Mutually parallel rotating electric and magnetic fields in the center of the disk give the maximum ME energy (or, in other words, the maximum helicity factor). Mutually perpendicular rotating electric and magnetic fields on the periphery of the disk give a power-flow vortex [55].

The spectral responses of a ferrite-disk ME meta-atom in a microwave waveguide and microwave cavity are defined by two external parameters – a bias magnetic field $H_0$ and a signal frequency $f$. The coherent quantum ME states are described with uncertainty in a bias magnetic field and frequency. When a magnetic field is suddenly changed, it introduces uncertainty in the energy of the magnetic system. Since energy and time are conjugate variables, a change in energy (due to the magnetic field change) will introduce an uncertainty in the time it takes for the system to adjust to the new field. This uncertainty in time corresponds to an uncertainty in the oscillation frequency. Since quantum transitions involve changes in energy levels, the uncertainty principle implies that the more precisely we know the energy change during a transition, the less precisely we can know the frequency.

Due to the *uncertainty principle*, fluctuations in quantum fields, existing in a very narrow frequency deviation $\Delta f$ and very narrow region of a bias magnetic field $\Delta H_0$, can be considered as virtual particles. Beyond the frames of the uncertainty limit, one has a continuum of energy. The spectra of ME oscillations in a quasi-2D ferrite disk are shown in Fig. 6. The sharply peaked (atomic-like) spectra from microwave experiments [59 – 61] may indicate macroscopic quantum phenomena with localized energy levels.

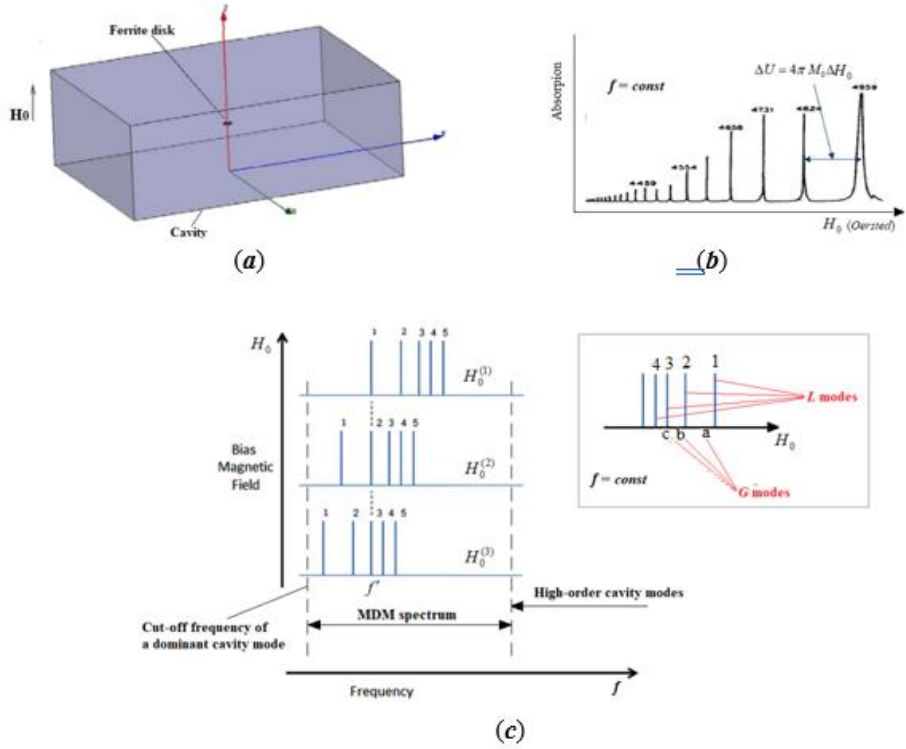

**Fig. 6.** Schematic representation of experimental resonances of a ferrite disk in a cavity. (***a***) microwave structure of a cavity with a ferrite disk. (***b***) Dillon's experimental results [59]. (***c***) A schematic representation of the sharply peaked (atomic-like) spectra from the microwave experiments. The entire spectrum lies in the frequency region where the high-order EM cavity modes are virtual photons. The inset shows the classification of the modes with respect to the bias magnetic field. *G* modes are dark modes in a spectrum (absorption lines). *L* modes are bright lines in a spectrum (emission lines). *G* modes are MDMs, while *L* modes are ME modes. The presented spectra show quantum transitions between *G* and *L* modes.

In a waveguide cavity used in microwave experiments, the frequency range of the entire spectrum lies above the cutoff frequency of a dominant mode and below cutoff frequencies of the high-order EM modes of the cavity [1]. It means that the *complete-set spectrum* of ME oscillations in a ferrite disk is viewed as *virtual photons* in the microwave cavity. In the spectra shown in Fig. 6, one observes quantum transitions of *G* and *L* modes. This refers to the changes in the quantum state of a system, involving the transfer of energy and symmetry properties between these modes. The *G*-mode to *L*-mode transitions are dynamical *symmetry breaking transitions*. This describes the situation where the rotationally symmetric *G* mode undergoes a change in its fundamental (ground) state due to internal dynamics, leading to a less symmetrical (excited) *L*-mode state.

Fig. 7 shows a dispersion characteristics of cavity EM modes and positions of *G* and *L* magnetic modes. Related to the dominant cavity mode, high-order cavity EM modes are complex (reactive) modes. Wave numbers of MDM oscillations (*G* modes) and ME oscillations (*L* modes) essentially exceed the wave numbers of EM oscillations in a microwave waveguide. The frequency $\omega$ of *G* mode oscillation is in the range between the frequencies $\omega_H = \mu_0 \gamma H_i$ and $\omega_\perp = \sqrt{\omega_H(\omega_H + \omega_M)}$. Here $\omega_M = \mu_0 \gamma M_0$, $\gamma$ is the gyromagnetic ratio, $H_i$ is a DC internal magnetic field, saturation magnetization $M_0$. In neglect of material anisotropy, the internal magnetic field is calculated as $H_i = H_0 - H_d$, where $H_0$ is a bias magnetic field and $H_d$ is a demagnetization field [39, 40]. The frequency of $L$ mode oscillation is around $2\omega$. The frequency range for $L$ mode oscillations is determined by the material and geometry parameters of the ferrite-disk sample. The range DC internal magnetic field for the $G$ and $L$ mode oscillations is $H_i^\diamond < H_i < \omega/\mu_0\gamma$, where $H_i^\diamond \equiv \sqrt{\left(\frac{\omega}{\mu_0\gamma}\right)^2 + \left(\frac{M_0}{2}\right)^2} - \frac{M_0}{2}$ [39, 40]. Interaction between $G$ modes and EM radiation is due to Zeeman effect. Interaction between $L$ modes and EM radiation is due to the relativistic effect of *rotational superradiance* [58].

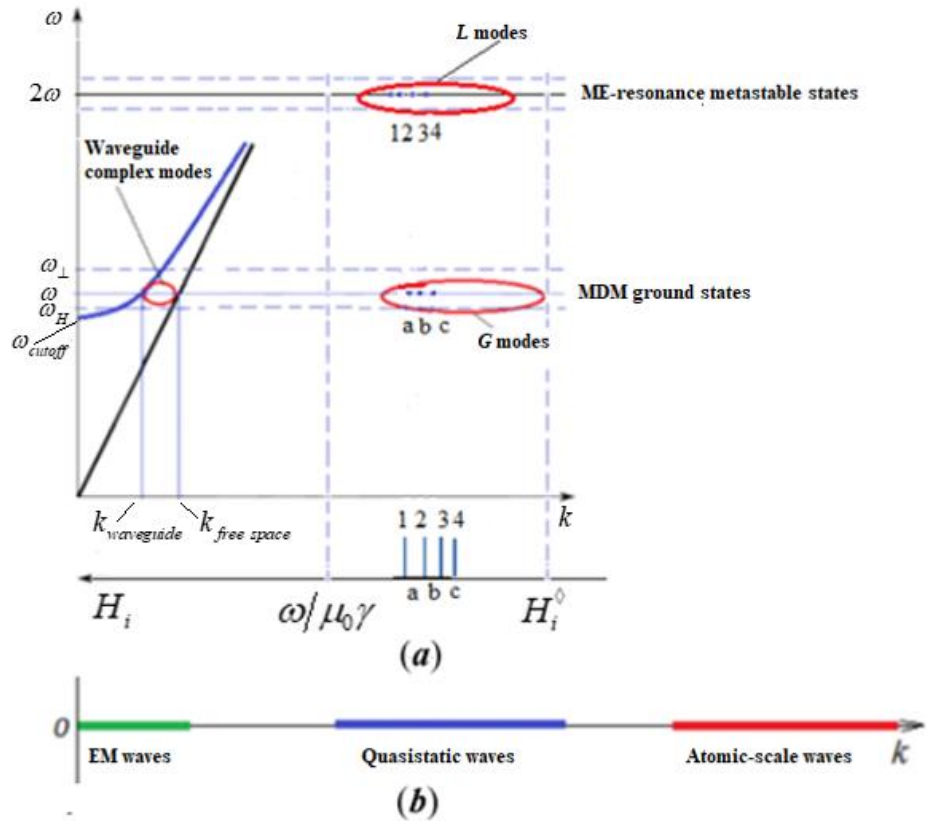

Fig. 7. A dispersion characteristic of cavity EM modes and positions of $G$ (designated by letters a, b, c...) and $L$ (designated by numbers 1, 2, 3...) magnetic modes. (*a*) Related to the dominant cavity mode, high-order EM cavity modes are complex (reactive) modes. Wave numbers of MDM oscillations ($G$ modes) and ME oscillations ($L$ modes) essentially exceed the wave numbers of EM oscillations in a microwave waveguide. (*b*) The $k$-space picture of the macroscopic, mesoscopic, and microscopic scales of oscillations.

Based on the above aspects, one can proceed to consider the problem of the ME quantum emitter. The ME quantum emission does not occur due to a combination of magnetic and electric dipole transitions. ME resonances are caused by the coupling effect between the MS and ES oscillations. This is a relativistic effect with symmetry breaking transitions. We have two-level quantum states. Each of these two-level states is a combination of quasistatic resonances with the cavity field. At one level, a combination of the MDM *G*-mode with a dominant linearly polarized cavity mode is observed. This is the Zeeman effect with macrospin structures. At another level, there is a combination of the ME *L*-mode with a *biorthogonal* cavity mode. This cavity mode is characterized by a topological field structure with a curved wave front. Metastable states can decay to lower energy states through magnetic dipole and electric quadrupole transitions.

The oscillation spectra of a microwave cavity with a ferrite disk are shown in Fig. 8. The resonances arise due to the interaction of two subsystems in a magnetic insulator. These are the subsystems of the ferromagnetic and electric polarization orders. The dynamical electric polarization is induced at magnetic resonances due to a topological effect. On the other hand, the electric polarization currents generate a topological contribution to the dynamical magnetization [43]. Bound states of two concurrent orders, caused by orbital angular momentum [44], are considered as ME states.

Magnetic energy of a sample in an external (bias) magnetic field is determined by the demagnetization field. The demagnetization field is the magnetic field generated by the magnetization in a magnet and the demagnetization factor determines how a magnetic sample responds to the bias magnetic field. It is evident that for the spectra, obtained at variation of a bias magnetic field $H_0$ and at a constant signal frequency $\omega$, a discrete reduction of magnetic energy of a ferrite disk should occur because of quantization of the demagnetization field. For an observer located on a ferrite disk, the magnetization dynamics is seen as the Larmor-clock magnon dynamics. There are standing-wave oscillations of *G*-mode MS wave functions. An observer located on a microwave cavity sees $2\omega$ rotations (*L*-mode magnon dynamics) of the wavefronts at the resonances. The *L*-mode magnetization dynamics of MS wave functions are accompanied by the electric polarization dynamics of ES wave functions. So, the *L* modes are ME modes.

The interaction of ferrite-disk MDM magnons with the EM cavity is via virtual photons. The balance of angular momenta of a ferrite disk and cavity walls is related by biorthogonal relations. A vacuum zone at the vicinity of a ferrite disk is a region determined by a complete-set spectrum of quasistatic oscillations. Because of the superradiance effect [58], the energy outside this zone will be greater than inside, resulting in a torque that rotates the cavity fields. Due to the uncertainty principle, virtual particles are considered as fluctuations in quantum fields that exist for a very narrow frequency deviation and very narrow region of a bias magnetic field. These virtual particles are created in pairs of "particles" and "anti-particles", which are the right-handed and left-handed helical modes. The variation of a bias magnetic field at a constant frequency of the microwave signal leads to the appearance of quantized edge magnetic and electric currents on the lateral surface of the ferrite disk and quantized electric and magnetic charges on the ferrite-disk planes. At the magnetic resonances, the cavity extracts energy from the internal energy of a ferrite disk. The cavity returns discrete portions of the energy to the microwave source. In other words, the cavity acquires *negative energy* relative to the microwave source feeding the cavity. One observes cavity negative-energy eigenstates at the ME resonances.

In Fig. 8, the oscillation spectrum of a microwave cavity with a ferrite disk, obtained at variation of a bias magnetic field $H_0$ at a constant signal frequency $\omega$, corresponds experimental results of Refs. [59 – 61]. The discretization of *internal* magnetic energy in a ferrite disk at $\omega$ = const is associated with the discretization of magnetization of the sample. When demonstrating the cavity negative-energy eigenstates at the magnetic resonances, it is assumed that the ground state of the ferrite *G*-modes corresponds to the "cavity level" – the reference level of the RF energy accumulated in the cavity at the maximum of the cavity resonance curve. An additional demonstration for the cavity transition for the second ME resonance peak in Fig. 9 clarifies this fact. Fig.10 shows quantum transitions of the internal magnetic energy of a ferrite disk in connection with transitions of discrete states of magnetization. Quantum transitions of cavity energy associated with magnetic transitions are shown in Fig. 11. The general picture of quantum levels of magnetic energy and cavity energy is shown in Fig. 12.

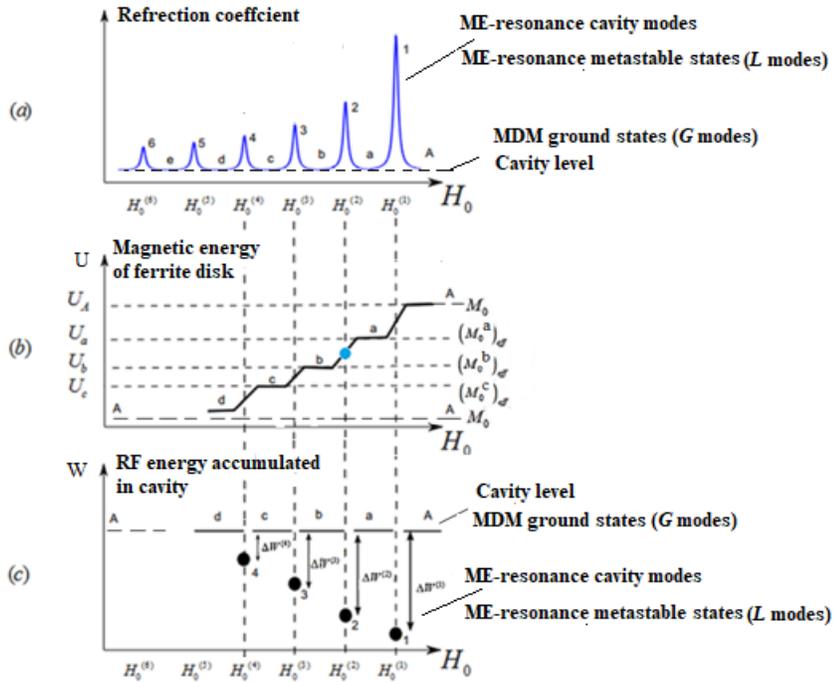

Fig. 8. (*a*) Theoretical model fo experimental oscillation spectra of a microwave cavity with a ferrite disk obtained at variation of a bias magnetic field $H_0$ at a constant signal frequency $\omega$ [59 – 61]. (*b*) Discretization of *internal* magnetic energy in a ferrite disk at $\omega$ = const is related to discretization of effective magnetization $\left(M_0^{(n)}\right)_{eff}$. (*c*) Cavity negative-energy eigenstates at the magnetic resonances. It is assumed that the "cavity level" corresponds to the ferrite *G*-mode ground state. The blue dot in Fig. 8 (*b*) indicates the second ME resonance peak.

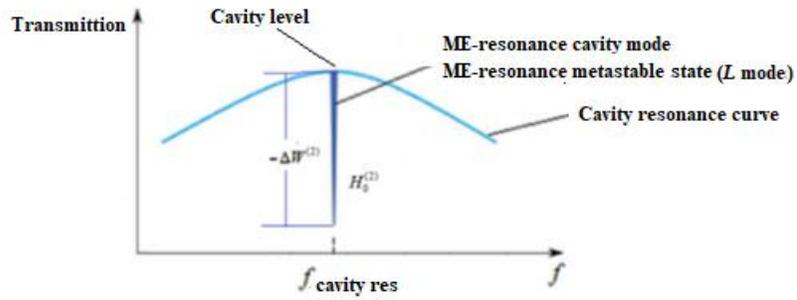

Fig. 9. The cavity transition coefficient for the second ME resonance peak.

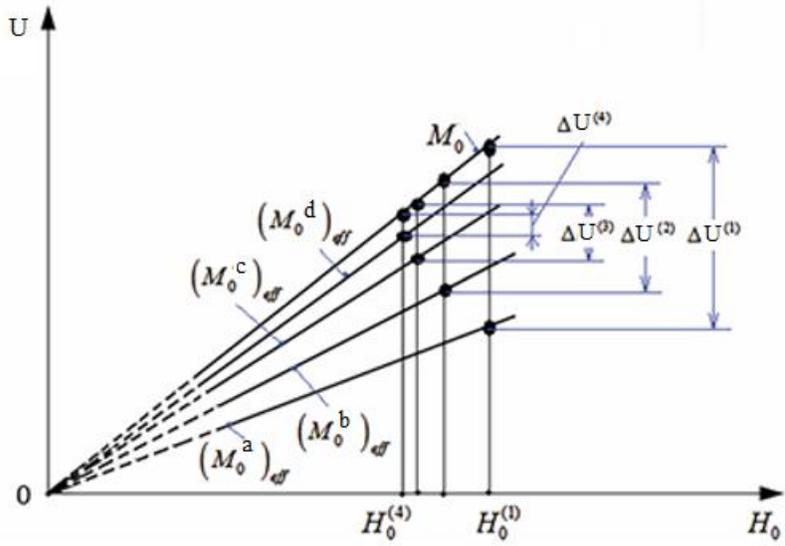

Fig.10. Quantum transitions of the internal magnetic energy of a ferrite disk in connection with transitions of discrete states of magnetization.

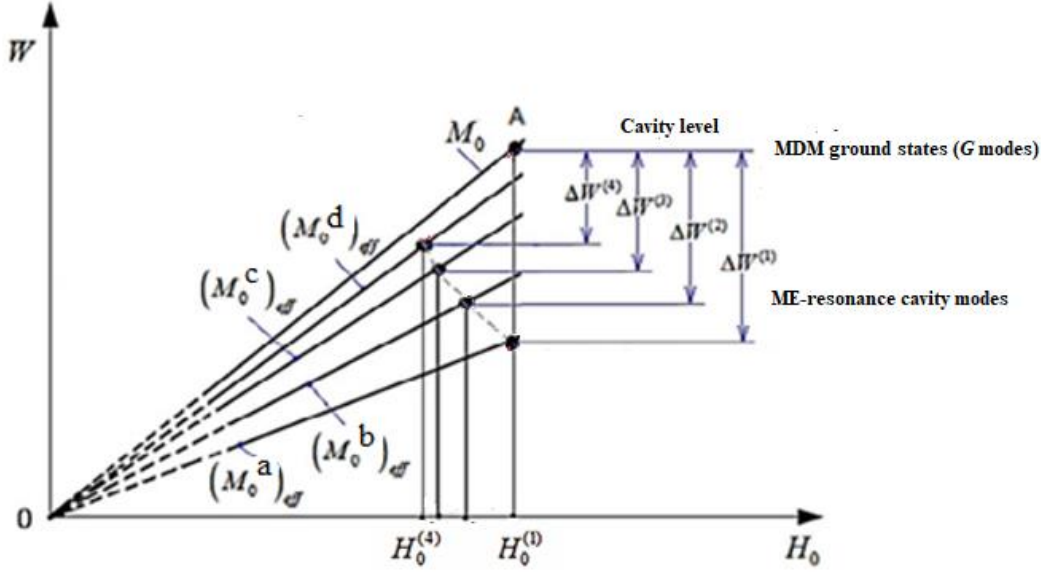

Fig. 11. Quantum transitions of cavity energy associated with transitions of discrete states of magnetization.

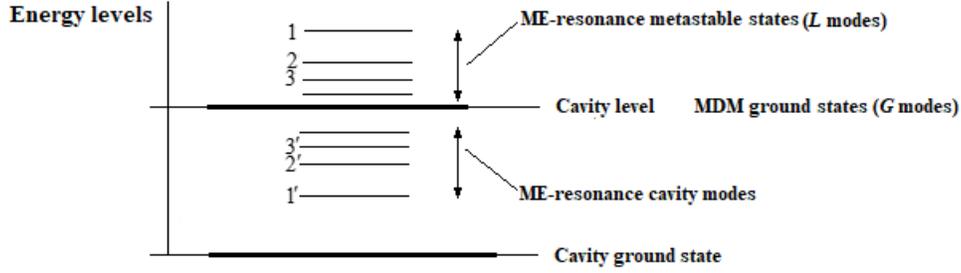

Fig. 12. The schematic of energy levels. By varying an external parameter (bias magnetic field), one can alter a quantum system's energy levels. These changes, observable as shifts in spectral lines, can be used to study symmetry breaking transitions.

## VI. HETEROCHIRAL HELICAL WAVE RESONANCES OF ME POLARITONS

The coupling of a subwavelength MDM resonator with a bosonic-field microwave cavity is far from a trivial problem. When a MDM ferrite disk is placed in a microwave cavity, one observes mixing space and time components, and the effects of spacetime curvature of the cavity fields. ME near fields emerge from the coupling of electric and magnetic dipole oscillations. These fields are characterized by a breaking of time-reversal and inversion symmetries and can exhibit features such as rotational superflows and quantized vortices. Hybrid states formed by the strong coupling of EM radiation in a cavity with ME excitations in a ferrite-disk meta-atom are ME polariton quantum states. In Ref. [58], we showed that magnon polaritons can be realized due to *magnon condensation*. The dipole–dipole condensed structures will be realizable when MDM magnons are strongly trapped by rotational superflow in a ring geometry of a quasi-2D ferrite disk.

This is considered as a main mechanism of strong coupling of the MDM with the microwave-cavity photon. Along with magnon condensation, we also have electric dipole condensation.

A proper understanding of the interaction between ferrite particles and an external EM field is based on an analysis of helical-mode quasistatic resonances, considered in Ref. [62]. It was shown that eigen states of oscillations in a quasi-2D ferrite disk are heterochiral helical resonances of scalar quasistatic wave functions. A heterochiral helix incorporates both right-handed and left-handed components. 3D heterochiral helical wave resonance can manifest characteristics resembling *double-frequency rotation on a 2D plane*. This involves projecting the 3D helical wave onto a 2D surface. The rotation of the fields in a ferrite disk is in the angular momentum balance with the rotation of currents and topological charges induced on the cavity walls. *ME virtual photons* are 3D heterochiral helical structures of quasistatic waves. The angular momentum (of total, spin and orbital) balances between the ferrite-disk and metal-wall planes are due to the ME virtual photons. Due to the topological action of the azimuthally unidirectional transport of energy in a ME-resonance ferrite sample, there exists the opposite topological reaction on a metal screen placed near this sample. We call this effect topological Lenz's effect [63]. The topological Lenz's law is applied to opposite topological charges: one in a ferrite sample and another on a metal screen. The MDM-originated near fields – the ME fields – induce helical surface electric currents and effective topological charges on the metal. The fields formed by these currents and charges will oppose their cause.

The Feigel hypothesis suggests that the quantum vacuum can transfer linear and angular momenta to the ME structure [28, 30]. Our situation is more complicated and interesting. When the *ME condensate rotates*, it interacts with vacuum quantum fluctuations in the cavity. This interaction causes the sample to emit ME photons, carrying away ME energy and angular momentum. The reaction force from the emission of these ME photons creates a torque on the ferrite disk. The angular momentum balance in a rotating polariton BEC involves a mechanism of balancing of the states. This balance is influenced by the system's broken rotational symmetry, which induces a net angular momentum and can lead to the formation of topological defects like vortices. The continuous inflow from an external system maintains the condensate's rotational state and prevents it from collapsing, enabling persistent currents. These currents, carrying angular momentum, can rotate continuously. As we noted above, the observed ME effect is extrinsic. This means that quantized rotating fields of $L$ modes with ME characteristics take place in the laboratory frame. In a quasi-rest frame of $G$ modes, such a ME effect is absent. When there is a phase shift $\pi$ in a quasi-rest frame of reference (for example, a phase shift of the Larmor- clock arrow), a phase shift $2\pi$ is observed in the laboratory frame of reference.

Fig. 13 illustrates some of the main aspects of the interaction of ME meta-atoms with the environment. Fig. 13 (*a*) shows how an interweaving of two types of rotating 3D heterochiral helical quasistatic modes, result in *double-frequency rotation* on a plane of a quasi-2D ferrite disk. For a disk placed in a close vicinity of two metal walls (it is assumed that there are no EM retardation effects between metal plates), we have 3D quasistatic solutions for heterochiral helical modes. This is shown in Fig. 13 (*b*) for the topological picture of the electric field distribution. The rotation of the electric field in a ferrite disk is in the angular momentum balance with the rotation of electric currents and topological charges induced on the metal walls. We can observe braided fields in vacuum cylinders above and below a ferrite disk. For a given direction of the bias magnetic field, the entire topological structure pattern rotates clockwise. The left insertion shows the numerical picture of the electric-field helical modes in a vacuum cylinder. The right insertion shows 3D heterochiral helical structures of the electric field inside a ferrite disk. The abbreviation TC in this insert denotes topological charges on the plane of the ferrite disk [58].

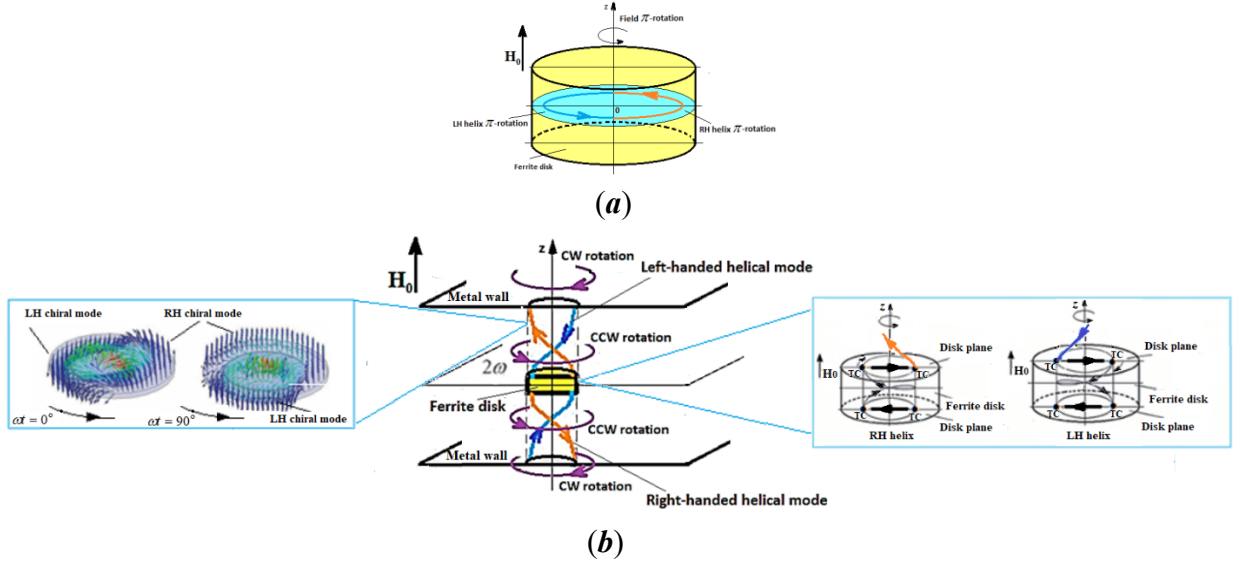

Fig. 13. Illustration of the result of interweaving of two types of rotating helical quasistatic modes on the $z = 0$ plane of a quasi-2D ferrite disk. 3D heterochiral helical wave can manifest characteristics resembling double-frequency rotation on a 2D plane. Aassuming the dynamic-phase $\pi$-rotation of the RF EM field (for example, a phase shift of the Larmor- clock arrow), we observe a coflowing turning over a regular-coordinate angle $2\pi$ due to $\pi$-rotation of the *RH* and *LH* helical qiasistatic modes [62]. The rotation of the electric field in a ferrite disk is in the angular momentum balance with the rotation of electric currents and topological charges induced on the metal walls. A heterochiral helical structure in vacuum incorporates both right-handed (red) and left-handed (blue) components.

In a "big" cavity, there are EM retardation effects between metal plates. We have an interaction between ME photons and "real" EM photons. The mechanism of coupling between ME photons and microwave photons is explained in Fig.14.

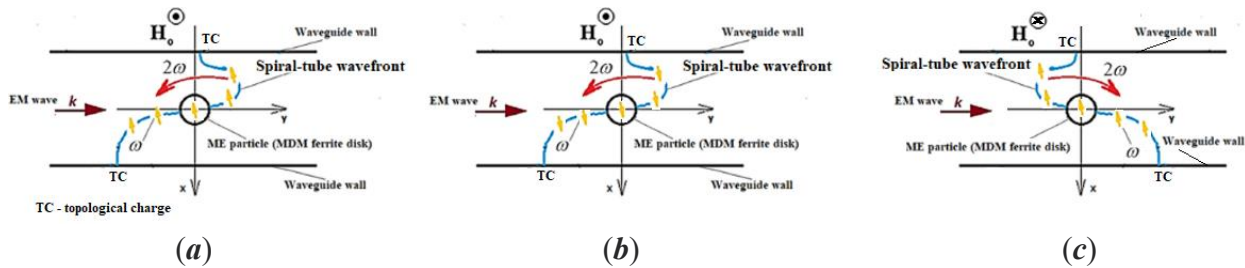

Fig. 14. Double-helix resonance of ME modes in a waveguide (cavity) at two directions of a bias magnetic field. The interaction of two perfectly matched rotating spiral waves creates a single structure: a "double spiral" with two arms. Wavefronts and topological-phase electric fields are shown at a given time phase, given directions of a bias magnetic field and given direction of EM-wave vector. Yellow arrows show the spatial orientations of the electric field vectors. The directions of these arrows coincide with the directions of the Larmor-clock arrows. There is synchronization between the arrows of Larmor clock of magnetization dynamics in the disk and arrows of the "clock" of surface electric currents on the metal walls [44]. (*a*) The field structure on a vacuum plane above a ferrite disk. (*b*) The field structure on a vacuum plane below a ferrite disk. (*c*) The field structure on a vacuum plane above a ferrite disk at the reversed direction of a bias magnetic field.

As can be seen in Fig. 14, ME vacuum states in a waveguide (cavity) are characterized by *P* and *T* symmetry breakings. We observe chiral radiation. The topological structure of the radiation depends on the direction of a bias magnetic field. It means that a subwavelength-sized ME particle can interact with the vacuum field, leading to the transfer of energy and momentum and the breaking of universal vacuum symmetries. Excited local ME emitters on an EM-wave front interact due to the increased coherence length of quasistatic oscillations. EM mode does not contribute to phase-matching considerations since EM propagation direction becomes inconsequential to the phase mismatch. We have spiral-tube structures of wave fronts. The coupling effect between the ME photon and the microwave-waveguide photon resembles a *rotary whirling sprinkler* which spins and sprays water in a circular pattern.

A ME polariton's structure, represented by a composition of three dispersion branches, is shown in Fig. 15. These are the microwave-waveguide photon branch, MDM branch, and ME-photon branch. The avoid-crossing interactions of these states result in hybrid states, which are polariton quantum states in a confined system.

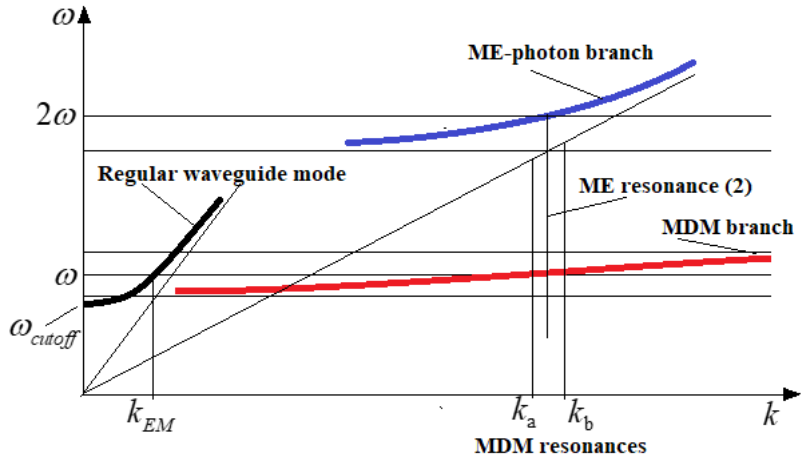

Fig. 15. The structure of the ME polariton, represented by a composition of three dispersion branches. The resonance peaks of Fig. 8 are shown as an example: the MDM peaks a and b, as well as the ME resonance peak 2.

## VII. DISCUSSION AND CONCLUSION

Our basic idea is that dynamic magnetoelectricity arises from magnetism similar to static magnetoelectricity. We should not regard true dynamic magnetoelectricity without taking into account the ME energy. We assume that the unification of EM and ME phenomena is essentially due to relativity. We introduce the concept of a ME meta-atom, a local source of the ME field. The ME meta-atom contains magnetic and dielectric subsystems. We consider the ME meta-atom as a quantum emitter with ME-energy eigenstates. The relativistic ME field has rotating electric- and magnetic-field components.

A subwavelength sample with ME oscillations – a ME meta-atom – has a strong influence on EM radiation. The near-field vacuum region of a ME meta-atom can be represented by ME virtual photons, which are heterochiral helical wave resonances. These virtual photons contribute to the electric, magnetic, and ME energies, as well as angular momenta in the vacuum. The fields of coupled ME-EM states have unique field topology with the $P$ and $T$ symmetry breaking effects. When ME virtual photons are mixed with actual EM photons, EM waves become split up for RH and LH helical modes of EM radiation. $P$ and $T$ symmetry breaking in spacetime refers to a scenario where the fundamental laws of physics do not hold true under mirror reflection and time reversal, or both. The ME meta-atom – can have a significant impact on the fundamental quantum behavior of vacuum. We can see that a local point with $P$ and $T$ symmetry breaking – the ME meta-atom – can lead to the symmetry breaking of the entire spacetime vacuum in cavity.

At the ME resonances, transfer between angular momenta in the magnetic insulator and in the vacuum cavity, demonstrates generation of vortex flows with fixed handedness. The unique topological properties of the polariton are manifested by curved wavefronts and the rotational supperradiance effects in microwave structures. In an environment of scattering states of a microwave cavity, EM waves carry topological phases of magnetic resonances in a ferrite disk. At the ME resonance peak (peaks 1, 2, 3…), the entire system, the cavity and disk, is characterized by electric energy $W_E$, magnetic energy $W_M$, and ME energy $W_{ME}$. The energy emitted from the cavity at the ME resonance peak is also characterized by three components of energy: $W_E$, $W_M$, and $W_{ME}$. The ME meta-atom creates a "pseudogravity" that deflects EM waves, similar to how actual gravitational fields do in general relativity.

To explain the interaction of ME meta-atoms with the environment, quantum field theory is needed. In view of the above discussions, we can formulate some basic aspects of ME quantum electrodynamics (ME QED) as follows. (*a*) ME meta-atoms are considered as 3D-confined subwavelength (mesoscopic) structures with interacting MS and ES resonances. These resonances are coupled due to relativistically rotating fields. (*b*) ME resonances in meta-atoms are characterized by energy eigenstates and have spin and orbital angular momenta. (*c*) ME meta-atoms have field structures called ME fields which are coherent states of virtual ME photons. Experimentally, these near-field ME photons appear in the context of a direct interaction between two closely located ME meta-atoms. The fields of virtual ME photons are characterized by both $P$ and $T$ symmetry breakings. (*d*) Along with the absorption and emission of virtual ME quanta, one can assume the existence of "real" ME photons, which are ME waves emitted by excited ME meta-atoms. Such ME waves (characterized by both $P$ and $T$ symmetry breakings) can be observed experimentally, when considering the interaction between two electromagnetically far-field distant ME meta-atoms in a system operating at a frequency below the cutoff frequency of the EM waveguide. In this case, the effective EM wavelength is stretched to infinity (the EM wave vector tends to zero). EM mode does not contribute any momentum to phase-matching considerations since EM propagation direction becomes inconsequential to the phase mismatch. This allows excited ME emitters to interact due to the increased coherence length. Because of relaxed phase matching conditions, this system provides enhancement of the fields of localized ME emitters, as well as the ME-field interference. (*e*) In the case when both the EM and ME wave propagation processes are involved, we will observe the effects of the ME-EM interaction.